\documentclass[a4paper,11pt]{article}
\usepackage[english]{babel}
\usepackage{mathrsfs}
\usepackage{amsfonts}
\usepackage[centertags]{amsmath}
\usepackage{amsthm}
\usepackage{enumitem}
\usepackage{latexsym,amsmath,amssymb}
\usepackage{graphicx,color}
\newcommand{\ds}{\displaystyle}

\setenumerate[1]{label=(\roman*), ref=(\roman*)}
\usepackage[all]{xy}

\newtheorem{thm}{Theorem}[section]
\newtheorem{prop}[thm]{Proposition}
\newtheorem{lemma}[thm]{Lemma}
\newtheorem{corol}[thm]{Corollary}

\newtheoremstyle{obs}
  {3pt}
  {3pt}
  {}
  {}
  {\bfseries}
  {.}
  {.5em}
  {}
\theoremstyle{obs}
\newtheorem{remark}[thm]{Remark}
\newtheorem{defn}[thm]{Definition}

\newcommand{\R}{\mathbb{R}}      
\newcommand{\F}{\mathbb{F}}
\def\qed{\ifvmode\removelastskip\fi
{\unskip\nobreak\hfil\penalty50\hbox{}\nobreak\hfil \hbox{\vrule
height1.2ex width1.2ex}\parfillskip=0pt \finalhyphendemerits=0
\par \smallskip}}

\title{Lagrangian submanifolds and Hamilton-Jacobi equation}

\author{\textsc{Mar\'ia Barbero-Li\~n\'an}\thanks{mbarbero@math.uc3m.es}\\
\small
Departamento de Matem\'aticas, Universidad Carlos III de Madrid, \\ \small Avenida de la Universidad 30, 28911 Legan\'es, Madrid, Spain
\\ \small and Instituto de Ciencias Matem\'aticas (CSIC-UAM-UC3M-UCM)\\ \and
\textsc{Manuel de Le\'on\thanks{mdeleon@icmat.es},}
\textsc{David Mart\'{\i}n de Diego}\thanks{david.martin@icmat.es} \\
\small
Instituto de Ciencias Matem\'aticas (CSIC-UAM-UC3M-UCM) \\ \small  C/Nicol\'as
Cabrera 13-15, 28049 Madrid, Spain
}

\begin{document}

\maketitle

\begin{abstract}
 Lagrangian submanifolds are becoming a very essential tool to generalize and geometrically understand
results and procedures in the area of mathematical physics. Here we use general Lagrangian submanifolds to
provide a geometric version of the Hamilton-Jacobi equation. This interpretation allows us to study
some interesting applications of Hamilton-Jacobi equation in holonomic, nonholonomic and time-dependent dynamics from a geometrical point of view.
\end{abstract}

\section{Introduction}

Hamilton-Jacobi procedure is useful to study the solution of  partial differential equations by analyzing the solutions of a system of ordinary differential equation and conversely.
Roughly speaking, the autonomous Hamilton-Jacobi problem for a Hamiltonian system on a configuration manifold $Q$ consists of finding
a function $S\colon Q\rightarrow \mathbb{R}$, $q\rightarrow S(q)$, such that
\begin{equation*}
\ds{H\left(q,\frac{\partial S}{\partial q} \right)=E},
 \label{eq:HJEqIntro}
\end{equation*}
where $H\colon T^*Q\rightarrow \mathbb{R}$ is the Hamiltonian function defined on the cotangent bundle $T^* Q$ and $E$ is a constant.

Recently, a new geometric approach to Hamilton-Jacobi theory has been developed~\cite{Cardin89, Cardin93, BcNZgzGHJ}. As a result, new developments
of Hamilton-Jacobi theory for nonholonomic mechanical systems~\cite{BCNZgz,Iglesias,AlmostPoisson} and in the discrete
setting~\cite{DiscreteHJ} have been obtained.

On the other hand, the notion of Lagrangian submanifolds, which are distinguished submanifolds of a symplectic manifold~\cite{Weinstein},
has gained a lot of interest due to its applications on dynamics~\cite{2012LJM,TuH,Tu}. These applications do not only appear on classical
dynamics, but also on more algebraic structures that generalize the notion of tangent bundle: the so-called Lie algebroids~\cite{LagranAlg}
and Lie affgebroids~\cite{LagrangAff}.

In this paper we show how general Lagrangian submanifolds are used to extend the geometric version of classical Hamilton-Jacobi equation along similar lines as in~\cite{Cardin89, Cardin93}. In~\cite{AbMa,LiMarle}, Lagrangian submanifolds obtained from a 1-closed form allowed to reinterpret the classical Hamilton-Jacobi
theorem in a more geometric way. Here, we consider a particular kind of Lagrangian submanifolds which allows us to describe Hamilton-Jacobi
equation for holonomic and nonholonomic dynamics. Moreover, we are also able to describe the time-dependent Hamilton-Jacobi equation in this geo\-me\-tric framework by means of a family of Lagrangian submanifolds of  $T^*Q$.
The main contribution of this paper is the application of only Lagrangian submanifold theory to obtain a geometric version of Hamilton-Jacobi equation in the above mentioned particular dynamics.

The paper is organized as follows. In Section~\ref{Sec:Lagrang}, we first review the notion of Lagrangian sub\-ma\-ni\-folds
and its characterizations. Section~\ref{Sec:ClassHJ} recalls the classical
Hamilton-Jacobi theory for both autonomous and non-autonomous cases. Section~\ref{Sec:GHJParticular}
contains the geometric interpretation of autonomous Hamilton-Jacobi equation for a general connected Lagrangian submanifold.
The definition of a family of Lagrangian submanifolds, more general than those being the image of a closed 1-form, allows us
to give  a geometric interpretation of Hamilton-Jacobi equation which is useful to describe Hamilton-Jacobi equation for holonomic and nonholonomic dynamics in Section~\ref{SubSec}. For the
nonholonomic case we need the language of Lagrangian distributions on symplectic
vector bundles to obtain the Hamilton-Jacobi equation, see Proposition~\ref{aqwe}. Another interesting contribution of this paper is the use of a family of
Lagrangian submanifolds to give a more generalized version of  the non-autonomous Hamilton-Jacobi equation in Section~\ref{Sec:TimeHJ}. As a result, this equation is described for holonomic and nonholonomic dynamics in Section~\ref{SubSecTimeHJ}.

\section{Lagrangian submanifolds}\label{Sec:Lagrang}

The essential background on symplectic geometry is reviewed here to make the paper as much self-contained as possible.
Lagrangian submanifolds~\cite{Weinstein} are the extension to manifolds of the notion of Lagrangian subspaces of
symplectic vector spaces.

Let us recall that a symplectic vector space is a pair $(E,\Omega)$ where $E$ is a vector space and $\Omega\colon E\times E \rightarrow \mathbb{R}$ is a skew-symmetric bilinear map of maximal rank. See~\cite{1989LeRo,GuiStern,LiMarle,Weinstein} for more details.

\begin{defn} Let $(E,\Omega)$ be a symplectic vector space and
$F\subset E$ a subspace. The $\Omega$-\textbf{orthogonal complement
of $F$} is the subspace defined by
\begin{equation*}
F^\perp=\{e\in E \; | \; \Omega(e,e')=0 \; \mbox{for all } e'\in
F\}.
\end{equation*}
The subspace $F$ is said to be
\begin{enumerate}
\item \textbf{isotropic} if $F\subseteq F^\perp$, that is,
$\Omega(e,e')=0$ for all $e,e'\in F$.
\item \textbf{Lagrangian} if $F$ is isotropic and has an
isotropic complement, that is, $E=F\oplus F'$, where $F'$ is
isotropic.
\end{enumerate} \label{defn:lagrangianSubspace}
\end{defn}

A well-known characterization of Lagrangian subspaces of finite dimensional symplectic vector spaces  is summarized in the following result:

\begin{prop} Let $(E,\Omega)$ be a finite dimensional symplectic vector space and
$F\subset E$ a subspace. Then the following assertions are
equivalent:
\begin{enumerate}
\item $F$ is Lagrangian,
\item $F=F^\perp$,
\item $F$ is isotropic and ${\rm dim} \, F=\frac{1}{2}{\rm dim}\,
E$.
\end{enumerate}\label{prop:LagrangianSubmanifoldIsotropic}
\end{prop}

As a consequence, we can characterize a Lagrangian subspace by checking if it has half the dimension of $E$ and $\Omega_{|F}=0$.

 Remember that a symplectic manifold
$(M,\omega)$ is defined by a differentiable manifold $M$ and a non-degenerate closed 2-form $\omega$ on $M$. Therefore, for each $x\in M$, $(T_xM, \omega_x)$ is a symplectic vector space
and a symplectic manifold has even dimension.

A symplectic vector bundle $(E,\omega,\pi)$ is a vector bundle $\pi\colon E\rightarrow M$ over a manifold $M$ equipped with a
smooth field $\omega$ of fiberwise nondegenerate skew-symmetric bilinear maps $\omega_x\colon E_x\times E_x \rightarrow \mathbb{R}$.
Therefore, if $(M,\omega)$ is a symplectic manifold, then $(TM,\omega,\tau_M)$ is a symplectic vector bundle, where
$\tau_M\colon TM \rightarrow M$ is the canonical tangent projection.
The converse is only true if
$\omega$ is closed.

The notion of Lagrangian subspace can be transferred to submanifolds by requesting that the tangent space of the
submanifold is a Lagrangian subspace for every point in the submanifold of a symplectic manifold.

\begin{defn} Let $(M,\omega)$ be a symplectic manifold and ${\rm
i}\colon N\rightarrow M$ an immersion. It is said that $N$ is an
\textbf{isotropic immersed submanifold} of $(M,\omega)$ if $({\rm
T}_x{\rm i})({\rm T}_xN) \subset {\rm T}_{{\rm i}(x)}M$ is an isotropic subspace
for each $x\in N$.  A submanifold $N\subset M$ is called
\textbf{Lagrangian} if it is isotropic and there is an isotropic
subbundle $P\subset TM|N$ such that $TM|N=TN\oplus
P$.\label{defn:lagrangianSubm}
\end{defn}

Note that ${\rm i}\colon N\rightarrow M$ is isotropic if and only if
${\rm i}^*\omega=0$, that is, $\omega({\rm T}_x{\rm i} (v_x), {\rm T}_x{\rm i} (u_x))=0$ for every $u_x,v_x\in T_xN$ and for every $x\in N$.

Of course, all the previous notions can be easily extended to the case of symplectic vector bundles. These extensions will be used in the sequel.

%

The canonical model of symplectic manifold is the cotangent bundle $T^*Q$ of an arbitrary manifold $Q$.
Denote by $\pi_Q\colon T^*Q \rightarrow Q$ the canonical projection and define a canonical 1-form $\theta_Q$ on $T^* Q$ by
\begin{equation*}
 \left(\theta_Q\right)_{\alpha_q}(X_{\alpha_q})=\langle \alpha_q, {\rm T}_{\alpha_q} \pi_Q(X_{\alpha_q})\rangle,
\end{equation*}
where $X_{\alpha_q}\in T_{\alpha_q}T^*Q$, $\alpha_q\in T^*Q$ and $q\in Q$. If we consider
bundle coordinates $(q^i,p_i)$ on $T^* Q$ such that $\pi_Q(q^i,p_i)=q^i$, then
\begin{equation*}\theta_Q=p_i  {\rm d}q^i\, .
 \end{equation*}
 The 2-form $\omega_Q=-{\rm d}\theta_Q$ is a symplectic form on $T^*Q$ with local expression
\begin{equation*}
\omega_Q={\rm d}q^i \wedge {\rm d}p_i.
\end{equation*}
The Darboux theorem states that this is the local model for an arbitrary symplectic manifold $(M,\omega)$: there exist local
coordinates $(q^i,p_i)$ in a neighbourhood of each point in $M$ such that $\omega={\rm d}q^i \wedge {\rm d}p_i$.

Note that the canonical 1-form $\theta_Q$ is universal in the sense that $\gamma^*(\theta_Q)=\gamma$ for an arbitrary 1-form $\gamma$ on $Q$. Hence
$\gamma^*(\omega_Q)=-{\rm d}\gamma$.

A relevant example of Lagrangian submanifold of the cotangent bundle is the following one.

\begin{prop} \cite{LiMarle} Let $\gamma$ be a 1-form on $Q$ and ${\mathcal L}=\hbox{Im } \gamma\subset T^*Q$.  The submanifold ${\mathcal L}$ of $T^*Q$ is Lagrangian if and only if
$\gamma$ is closed. \label{prop:LagrangianClosedForm}
\end{prop}
The result follows because $\dim {\mathcal L}=\dim Q$ and $\gamma^*(\omega_Q)=-{\rm d}\gamma$.

Given  a symplectic manifold $(P, \omega)$, $\dim P=2n$, it is well-known that its tangent bundle $TP$ is equipped with a
symplectic structure denoted by $\mathrm{d}_T\omega$, where ${\rm d}_T \omega$ denotes the tangent lift of $\omega$ to $TP$. If we
take Darboux coordinates $(q^i,p_i)$ on $P$, that is, $\omega=\mathrm{d}q^i\wedge \mathrm{d}p_i$, then
$\mathrm{d}_T\omega=\mathrm{d}\dot{q}^i\wedge \mathrm{d}p_i+\mathrm{d}q^i\wedge \mathrm{d}\dot{p}_i$, where
$(q^i, p_i; \dot{q}^i, \dot{p}_i)$ are the induced coordinates on $TP$. We will denote the bundle coordinates on $T^*P$ by
$(q^i, p_i; a_i, b^i)$, then $\omega_{P}=\mathrm{d}q^i\wedge \mathrm{d}a_i+\mathrm{d}p_i\wedge \mathrm{d}b^i$.
If we denote by $\flat: TP\to T^*P$ the isomorphism defined by $\omega$, that is, $\flat_{\omega}(v)={\rm i}_{v}\,\omega$, then
we have $\flat_{\omega}(q^i, p_i; \dot{q}^i, \dot{p}_i)=(q^i, p_i; -\dot{p}_i, \dot{q}^i)$.

Given a function $H: P\to \R$, and its associated Hamiltonian vector field $X_H$, that is, ${\rm i}_{X_H}\omega=\mathrm{d}H$,
then the image of  $X_H$, $\hbox{Im } X_H$, is a Lagrangian submanifold of  $(TP, \mathrm{d}_T\omega)$.

In Sections \ref{Sec:GHJParticular} and \ref{Sec:TimeHJ} we will show other examples of Lagrangian submanifolds (see also \cite{LiMarle}) that play a key role to develop the novel applications of Lagrangian submanifolds on Hamilton-Jacobi theory described in this paper.

\section{Classical Hamilton-Jacobi theory}\label{Sec:ClassHJ}

Once we have introduced the necessary background on symplectic geometry, let us review briefly Hamiltonian dynamics.  If $H\colon T^*Q \rightarrow \mathbb{R}$
is a Hamiltonian function, the Hamiltonian vector field $X_H$ on $T^* Q$ is the solution of the equation
\begin{equation*}
{\rm i}_{X_H} {\omega_Q}={\rm d}H.
\end{equation*}
If we
take Darboux coordinates $(q^i,p_i)$ on $T^*Q$, Hamilton's equations are locally
written as
\begin{eqnarray*}
 \dfrac{{\rm d} q^i}{{\rm d} t}&=& \dfrac{\partial H}{\partial p_i}(q,p),\\
 \dfrac{{\rm d} p_i}{{\rm d} t}&=& -\dfrac{\partial H}{\partial q^i}(q,p).
\end{eqnarray*}

Hamilton-Jacobi theory gives us useful tools to integrate some particular solutions of Hamilton's equation, though we will not
always find a solution. Let us review here the classical  Hamilton-Jacobi theory both in the autonomous and non-autonomous cases.

The following is a geometric version of the classical Hamilton-Jacobi theory~\cite{AbMa,Arnold} in the autonomous
case.

\begin{thm}[Hamilton-Jacobi equation \cite{AbMa}] Let $T^*Q$ be the symplectic manifold
with the symplectic structure $\omega_Q=-{\rm d} \theta_Q$. Let $X_H$ be
a Hamiltonian vector field on $T^*Q$, and let $S\colon
Q\rightarrow \mathbb{R}$ be a smooth function. The following two conditions are
equivalent:
\begin{enumerate}
\item for every curve $c$ in $Q$ satisfying
\begin{equation*} c'(t)={\rm T}\pi_Q (X_H({\rm d}S(c(t))))
\end{equation*}
the curve $t\mapsto {\rm d} S(c(t))$ is an integral curve of $X_H$.
\item $S$ satisfies the Hamilton-Jacobi equation $H\circ {\rm d}
S=E$, a constant, that is,
\begin{equation*}
H\left(q,\frac{\partial S}{\partial q}\right)=E.
\end{equation*}
\end{enumerate} \label{thm:HJ}
\end{thm}

\remark \label{RmkHJ} The above theorem describes solutions to Hamilton-Jacobi equation by finding particular integral curves
of the Hamiltonian vector field. Those integral curves are obtained, when possible, from integral curves of a vector
field on $Q$ defined by $X^{{\rm d}S}={\rm T}\pi_Q \circ X_H \circ {\rm d}S$. In other words, the following diagram commutes for
a function $S$ on $Q$:
\begin{equation*}\xymatrix{ T^*Q \ar@/^1pc/[d]^{\txt{\small{$\pi_Q$}}} \ar[r]^{\txt{\small{$X_H$}}}&  TT^*Q\ar[d]^{\txt{\small{${\rm T}\pi_Q$}}}
\\
Q   \ar@/^1pc/[u]^{\txt{\small{${\rm d} S$}}} \ar[r]^{\txt{\small{$X^{{\rm d}S}$}}} & TQ}
\end{equation*}
The integral curves $c$ of $X^{{\rm d}S}$ define the integral curves ${\rm d}S\circ c$ of the Hamiltonian vector field $X_H$.

 Moreover, note that the set $N=\{(q,{\rm d}S(q)) \; | \; q\in Q\}$ of $T^*Q$ is a Lagrangian submanifold according
to Proposition~\ref{prop:LagrangianClosedForm} since it is defined by a closed 1-form, ${\rm d}S \colon Q\rightarrow T^*Q$.
Thus the notion of Lagrangian submanifolds already appears in the classical version of Hamilton-Jacobi equation.
 Then it makes sense to look for an extension of this result to any Lagrangian submanifold as shown in Section~\ref{Sec:GHJParticular}
and consider the applications, for instance, to holonomic and nonholonomic dynamics, see Section~\ref{SubSec}.

The classical statement of Hamiton-Jacobi equation for non-autonomous case is the following:
\begin{thm}[Time-dependent Hamilton-Jacobi equation \cite{AbMa}] Let $T^*Q$ be the symplectic manifold
with the symplectic structure $\omega_Q=-{\rm d} \theta_Q$. Let
$X_H$ be a Hamiltonian vector field on $T^*Q$ and  $W\colon \mathbb{R}\times Q\rightarrow \mathbb{R}$ be a smooth function.
The following two conditions are equivalent:
\begin{enumerate}
\item for every curve $c$ in $Q$ satisfying
\begin{equation*} c'(t)={\rm T}\pi_Q (X_{H}({\rm d}W_t(c(t))))
\end{equation*}
the curve $t\mapsto {\rm d} W_t(c(t))$ is an integral curve of
$X_H$, where $W_t\colon Q\rightarrow \mathbb{R}$, $W_t(q)=W(t,q)$.
\item $W$ satisfies the Hamilton-Jacobi equation
\begin{equation*}
H\left(q,\dfrac{\partial W}{\partial q}\right)+\dfrac{\partial W}{\partial t}= \hbox{
constant on} \;\; T^*Q,
\end{equation*}
that is, $
\ds{H\circ {\rm d}W_t +\dfrac{\partial W}{\partial t}= K(t)}$.
\end{enumerate} \label{thm:HJtime}
\end{thm}

\remark \label{RmkHJTime} The analogous diagram to the one in Remark~\ref{RmkHJ} for the time-dependent case is the following one:
\begin{equation*}\xymatrix{ T^*Q \ar@/^1pc/[d]^{\txt{\small{$\pi_Q$}}} \ar[r]^{\txt{\small{$X_{H}$}}}&  TT^*Q\ar[d]^{\txt{\small{${\rm T}\pi_Q$}}}
\\
Q   \ar@/^1pc/[u]^{\txt{\small{${\rm d} W_t$}}} \ar[r]^{\txt{\small{$X^{{\rm d}W_t}$}}} & TQ}
\end{equation*}
which must be understood as in Remark~\ref{RmkHJ} for every time $t$.

In the non-autonomous case the set $N_t=\{(q,{\rm d}W_t(q)) \; | \; q\in Q\}$ is also a
Lagrangian submanifold of $T^*Q$ for every $t$. In Section~\ref{Sec:TimeHJ} we will discuss the time-dependent
Hamilton-Jacobi theorem in an intrinsic way for a general family of Lagrangian submanifolds and a suitable family of submanifolds in $TT^*Q$
so that everything comes together.

\section{A geometric version  of Hamilton-Jacobi equation}\label{Sec:GHJParticular}

According to the philosophy of this paper we use arbitrary Lagrangian submanifolds ${\mathcal L}$ of the cotangent bundle $T^*Q$  to extend geometrically Theorem \ref{thm:HJ}. To be more precise we consider
a Lagrangian submanifold ${\mathcal L}$ in $T^*Q$ that does not necessarily
project over the entire manifold $Q$ as in the case of Theorem \ref{thm:HJ}, see Remark~\ref{RmkHJ}. Let ${\rm i}_{\mathcal L}\colon {\mathcal L} \rightarrow T^*Q$ be
the canonical
immersion, the canonical tangent lift of ${\rm i}_{\mathcal L}$ is denoted by ${\rm T}{\rm i}_{\mathcal L}\colon T{\mathcal L} \rightarrow TT^*Q$.

It is also possible to define the canonical lift of ${\rm i}_{\mathcal L}$ to the cotangent bundle, but only on the points $p_q\in {\mathcal L}$.
 Let $\langle \cdot ,\cdot \rangle$
be the inner product, then the canonical lift of ${\rm i}_{\mathcal L}$ to the cotangent bundle, ${\rm T}^*{\rm i}_{\mathcal L}\colon T^*_{\mathcal L}T^*Q\rightarrow T^*{\mathcal L}$, is
defined as follows
\begin{equation} \langle{\rm T}^*_{{\rm i}_{\mathcal L}(p_q)}{\rm i}_{\mathcal L}(\Lambda_{p_q}),
X_{p_q}\rangle= \langle \Lambda_{p_q},
{\rm T}_{p_q}{\rm i}_{\mathcal L}\left(X_{p_q}\right)\rangle, \label{Eq:PullbackI} \end{equation}
for $p_q\in {\mathcal L}$, $\Lambda_{p_q}\in T^*_{\mathcal L}T^*Q$,
$X_{p_q}\in T_{p_q}{\mathcal L}$. This notion corresponds to the pull-back of ${\rm i}_{\mathcal L}$ induced on differential forms.

The situation can be summarized in the following diagram:
\begin{equation*}\xymatrix{ T{\mathcal L}\; \ar@{^{(}->}[rr]^{\txt{\small{$ {\rm T} {\rm i}_{\mathcal L}$}}} \ar[ddrr]_{\txt{\small{$\tau_{\mathcal L}$}}} && TT^*Q \ar@/^1pc/[dd]^{\txt{\small{$\tau_{T^*Q}$}}} \ar[rr]^{\txt{\small{$\flat$}}}
&& T^*T^*Q\; \ar@/^1pc/[dd]^{\txt{\small{$\pi_{T^*Q}$}}}
\ar@{^{(}->}[rr]^{\txt{\small{$ {\rm T}^* {\rm i}_{\mathcal L}$}}} && T^* {\mathcal L}   \ar[ddll]^{\txt{\small{$\pi_{\mathcal L}$}}}
\\ \\
 &&{\mathcal L} \ar@{=}[rr]  \ar@/^1pc/[uu]^{\txt{\small{$\left.X_H\right|_{\mathcal L}$}}} && {\mathcal L}
 \ar@/^1pc/[uu]^{\txt{\small{$\left.{\rm d} H\right|_{\mathcal L}$}}} && }
\end{equation*}

Here, the map $\flat$ is the symplectic musical isomorphism with inverse given by $\sharp\colon T^*T^*Q\rightarrow TT^*Q$. Moreover note that in the above diagram we have identified ${\mathcal L}$ with its image by ${\rm i}_{\mathcal L}$.
Now  we can generalize Theorem~\ref{thm:HJ} to any Lagrangian submanifold of $T^*Q$.

\begin{thm} Let ${\mathcal L}$ be a  Lagrangian submanifold of $T^*Q$. The following statements are equivalent:
\begin{enumerate}
\item ${\rm T}^*{\rm i}_{\mathcal L}\left(\left.{\rm d}H\right|_{\mathcal L}\right)=0$, where
${\rm i}_{{\mathcal L}}\colon {\mathcal L}
\hookrightarrow T^*Q$ is the canonical immersion;
\item ${\rm Im} \left(({\rm d}H)_{|\mathcal L}  \right)
\subset (T{\mathcal L})^0$, where $ (T{\mathcal L})^0$ is the annihilator of $T{\mathcal L}$;
\item ${\rm Im} \left( (X_H)_{|{\mathcal L}}\right)
\subset T{\mathcal L}$.
\end{enumerate}

\proof
Note that condition $(i)$, ${\rm T}^*{\rm i}_{\mathcal L}\left(\left.{\rm d}H\right|_{\mathcal L}\right)=0$, is equivalent to $H|_{{\mathcal L}}$ being  constant  and it is directly equivalent to condition $(ii)$.
Now, let $\sharp \colon T^*T^*Q \rightarrow TT^*Q$ be the symplectic musical isomorphism, then the equivalence between $(ii)$ and $(iii)$ is based on
the following equivalence:
\begin{equation*}
\sharp (({\rm d}H)_{|\mathcal L} )\subset \sharp \left((T{\mathcal L})^0\right) \quad \Leftrightarrow  \quad
X_H(p_q) \subset \left( T_{p_q}{\mathcal L}\right)^\perp=T_{p_q}{\mathcal L} \quad \forall \quad p_q\in {\mathcal L},
\end{equation*}
where we first have used that $\ds{\left({\rm T} {\mathcal L}\right)^\perp=
\sharp\left(\left({\rm T} {\mathcal L}\right)^0\right)}$ by~\cite[Proposition 13.2]{LiMarle}.
Observe that we have only used the Lagrangian character of ${\mathcal L}$ in the last step because $ \left( T_{p_q}{\mathcal L}\right)^\perp=T_{p_q}{\mathcal L} $ for every $p_q\in {\mathcal L}$.
\qed \label{Prop:ExtensionHJ}
\end{thm}

Let us rewrite the above theorem in local coordinates. If ${\mathcal L}$ is locally given by the independent constraints $\Phi^i(q,p)=0$ for $i=1,\dots, \dim Q$, then the condition of being Lagrangian is written as
\[
\{\Phi^i, \Phi^j\}=\frac{\partial\Phi^i}{\partial q^k}\frac{\partial\Phi^j}{\partial p_k}-\frac{\partial\Phi^i}{\partial p_k}\frac{\partial\Phi^j}{\partial q^k}=0, \quad \forall \quad i, j=1,\dots,{\rm dim}Q\; .
\]
where $\{\, \cdot \,  ,\,\cdot \, \}$ is the canonical Poisson bracket of functions on $T^*Q$ induced by ${\omega_Q}$ .

The tangent bundle of ${\mathcal L}$ is described locally as follows:
\begin{eqnarray*}
T{\mathcal L}&=&\left\{(q,p,\dot{q},\dot{p})\in TT^*Q \, \Big| \, \Phi^i(q,p)=0, \; \dfrac{\partial \Phi^i}{\partial q^j} \dot{q}^j +
\dfrac{\partial \Phi^i}{\partial p_j} \dot{p}_j=0 \right\} \\
&=&\left\{(q,p,\dot{q},\dot{p})\in TT^*Q \, \Big| \, \Phi^i(q,p)=0, \; \dot{p}_j=-\lambda_k \dfrac{\partial \Phi^k}{\partial q^j},  \dot{q}^j=\lambda_k
\dfrac{\partial \Phi^k}{\partial p_j}  \right\},
\end{eqnarray*}
where $\lambda_k$ are functions on ${\mathcal L}$.

Moreover,
\begin{equation*}
\left(T{\mathcal L}\right)^0=\left\{(q,p,\alpha,\beta) \in T^*T^*Q \, \Big| \, \Phi^i(q,p)=0, \; \alpha_i=\lambda_k \dfrac{\partial \Phi^k}{\partial q^i}, \; \beta_i=\lambda_k \dfrac{\partial \Phi^k}{\partial p_i}
 \right\}.
\end{equation*}

The standard case for Lagrangian submanifolds is given by a closed  1-form   $\gamma$ on $Q$ as mentioned in  Proposition~\ref{prop:LagrangianClosedForm}.
 Observe that  Theorem  \ref{Prop:ExtensionHJ} is an extension of  Theorem \ref{thm:HJ} as mentioned in Remark~\ref
 {RmkHJ} considering  ${\mathcal L}={\rm Im} \, {\rm d}S$, where $S\colon Q \rightarrow \mathbb{R}$.
Let us define a vector field $X_H^{{\rm d}S}$ on $Q$ by
\[
X_H^{{\rm d}S}=  {\rm T}\pi_Q\circ  X_H\circ {\rm d}S.
\]
Since ${\rm d}S$ is a diffeomorphism onto its image ${\mathcal L}={\rm Im} \, {\rm d}S$,
it is equivalent to find the integral curves of  $X_H^{{\rm d}S}$ and to find the integral curves of the vector field $X_H|_{{\mathcal L}}\in {\mathfrak X} ({\mathcal L})$, see~\cite{BcNZgzGHJ}.

Another interesting example of Lagrangian submanifolds is given  by a
submanifold $\mathcal{M}$ immersed in $Q$, ${\rm i}_{\mathcal{M}}\colon
\mathcal{M} \hookrightarrow Q$, and a closed 1-form  $\gamma$ on $M$,
as follows
\begin{equation}
{\mathcal L}_{\mathcal{M},\gamma} =\{\mu \in T^*Q \; | \;
{\rm T}^*{\rm i}_{\mathcal{M}}(\mu)=\gamma\} =\{\mu \in T^*Q \; | \; \langle \mu_{{\rm i}_{\mathcal M}(x)}, {\rm T}_x{\rm i}_{\mathcal M}(v_x)\rangle
= \langle \gamma_x, v_x \rangle  \; \forall \; v_x\in T_x{\mathcal M} \},
\label{eq:Sigma_mf}
\end{equation}
where ${\rm T}^*{\rm i}_{\mathcal{M}}$ has been defined in~\eqref{Eq:PullbackI}.

Note that locally, $\gamma={\rm d}f$ where $f\in C^{\infty}({\mathcal M})$.
\begin{lemma} Let $\gamma$ be a closed 1-form on ${\mathcal M}$.
 The submanifold ${\mathcal L}_{\mathcal{M},\gamma} =\{\mu \in T^*Q \; | \;
{\rm T}^*{\rm i}_{\mathcal{M}}(\mu)=\gamma \}$ is a Lagrangian submanifold of $T^*Q$.
\proof It can be proved straightforward by using local coordinates and the closedness of $\gamma$. Assume that locally
${\rm i}_{\mathcal M}(q^a)=(q^a,\psi^\alpha(q^a))$, then
\begin{equation*}
 {\mathcal L}_{\mathcal{M},\gamma} =\left\{(q^a,q^\alpha,\mu_a,\mu_\alpha) \in T^*Q \, | \, q^\alpha=\psi^\alpha(q^a), \;
\mu_a=\gamma_a-\dfrac{\partial \Psi^\alpha}{\partial q^a} \mu_\alpha \right\}. \label{eq:LocalLMgamma}
\end{equation*}
Obviously $\dim \,  {\mathcal L}_{\mathcal{M},\gamma}= \dfrac{1}{2} \dim \, T^*Q$. It remains to prove that
${\rm T}^*{\rm i}_{{\mathcal L}_{\mathcal{M},\gamma}}({\omega_Q})=0$, that is, \begin{equation*}{\omega_Q}\left( {\rm T}{\rm i}_{{\mathcal L}_{\mathcal{M},\gamma}}(X_{\mu_q}),
{\rm T}{\rm i}_{{\mathcal L}_{\mathcal{M},\gamma}}(Y_{\mu_q}) \right)=0 \quad \mbox{for all } X_{\mu_q}, \, Y_{\mu_q}\in T_{\mu_q}{\mathcal L}_{\mathcal{M},\gamma}.
\end{equation*}
\qed \label{Lemma:SigmaNf}
\end{lemma}

\section{Applications of time-independent Hamilton-Jacobi theory} \label{SubSec}

In this section we use those Lagrangian submanifolds defined in~(\ref{eq:Sigma_mf}) to obtain
some interesting and novel applications of Theorem~\ref{Prop:ExtensionHJ} in Hamilton-Jacobi theory for the autonomous case.

\subsection{Holonomic dynamics} Let $N$ be a submanifold of $Q$ and $h$ be a Hamiltonian function on $T^*N$. Take a closed 1-form $\gamma$ on $N$.
Then, we have that $\hbox{Im}\, \gamma$ is a Lagrangian submanifold of $T^*N$ and at the same time
we can define a Lagrangian submanifold  ${\mathcal L}_{N,\gamma}$ of $T^*Q$ according to~\eqref{eq:Sigma_mf} which  is given by
\begin{equation*}
{\mathcal L}_{N,\gamma}=\{\mu\in T^*Q \; | \; {\rm T}^*{\rm i}_N(\mu)=\gamma\}=\{\mu \in T^*Q \; | \; \mu_{{\rm i}_N(x)}({\rm T}_x{\rm i}_N(v_x))=\gamma_x(v_x) \; \forall \; v_x\in T_xN \},
\end{equation*}
where ${\rm i}_N \colon N \hookrightarrow Q$ is the canonical immersion. The following diagram illustrates the above situation
\begin{equation*}
 \xymatrix{ \mathbb{R} & T^*N \ar[l]_{\txt{\small{$h$}}} \ar[dr]^{\txt{\small{$\pi_N$}}} &
T^*Q \ar[d]^{\txt{\small{$\left.\pi_Q\right|_{{\mathcal L}_{N,\gamma} }$}}} \ar[l]_{\txt{\small{${\rm T}^*{\rm i}_{N}$}}}  & {\mathcal L}_{N,\gamma}
 \ar@{_{(}->}[l]_{\txt{\small{${\rm i}_{{\mathcal L}_{N,\gamma}}$}}} \ar[d] \\ & & N  \ar@{=}[r]  & N
}
\end{equation*}
Hence we can extend the Hamiltonian $h$ to a hamiltonian $H: T^*_N Q\longrightarrow {\mathbb R}$ just defining $H= h\circ {\rm T}^*{\rm i}_N$.

All the elements in Theorem~\ref{Prop:ExtensionHJ} for the Lagrangian submanifold  ${\mathcal L}_{N,\gamma}$ are gathered in the fo\-llo\-wing diagram:
\begin{equation*}
 \xymatrix{ \hbox{Im} \left(\left. X_H \right|_{{\mathcal L}_{N,\gamma}} \right)\ar@{^{(}->}[dr] & & &
\hbox{Im}\left(\left. {\rm d}H \right|_{{\mathcal L}_{N,\gamma}} \right)\ar@{_{(}->}[dl]\\
  {\rm T} {\mathcal L}_{N,\gamma} \ar[dr]^{\txt{\small{$\tau_{{\mathcal L}_{N,\gamma}} $}}} \ar@{^{(}->}[r]^{\txt{\small{$T{\rm i}_{{\mathcal L}_{N,\gamma}} $}}}&T_{{\mathcal L}_{N,\gamma}}T^*Q
\ar[d]^{\txt{\small{$\tau_{T^*Q}$}}} \ar@/^1pc/[r]^{\txt{\small{$\flat$}}} & T^*_{{\mathcal L}_{N,\gamma}}T^*Q
\ar[d]^{\txt{\small{$\pi_{T^*Q}$}}} \ar@/^1pc/[l]_{\txt{\small{$\sharp$}}} &
({\rm T} {\mathcal L}_{N,\gamma})^0   \ar@{_{(}->}[l] \ar[dl] \\ & {\mathcal L}_{N,\gamma} \ar[d]^{\txt{\small{$\pi$}}}  \ar@{=}[r]  & {\mathcal L}_{N,\gamma} \ar[d]^{\txt{\small{$\pi$}}} & \\
& N \ar@{=}[r]  &  N & }
\label{Diagram:HJHolonomic}
\end{equation*}

\begin{prop}\label{holonomic} The standard Hamilton-Jacobi equation for holonomic dynamics on $N$
\[ \gamma^* {\rm d}h=0, \mbox{ that is }, h\circ \gamma= {\rm constant},
\]
is equivalent to
\[
{\rm T}^* {\rm i}_{{\mathcal L}_{N,\gamma}}\left({\rm d}H\right)=0,
\]
where $H: T^*_N Q\longrightarrow {\mathbb R}$, $H= h\circ {\rm T}^*{\rm i}_N$.
\proof
Let us start from ${\rm T}^* {\rm i}_{{\mathcal L}_{N,\gamma}}\left({\rm d}H\right)=0$ and use the definition of the Hamiltonian function
$H$ on $T^*_NQ$. For $\mu_q\in {\mathcal L}_{N,\gamma}$,
\begin{eqnarray*}
 0&=& {\rm T}^* {\rm i}_{{\mathcal L}_{N,\gamma}}\left({\rm d}H (\mu_q)\right)\\
&=&{\rm T}^* {\rm i}_{{\mathcal L}_{N,\gamma}}\left({\rm d}\left(h\circ {\rm T}^*{\rm i}_N\right) (\mu_q)\right)\\
&=&{\rm T}^* {\rm i}_{{\mathcal L}_{N,\gamma}}\left({\rm d}\left(h\circ \gamma (\tilde{q})\right)\right)\\
&=&{\rm T}^* {\rm i}_{{\mathcal L}_{N,\gamma}}\left( \gamma^* {\rm d}h(\tilde{q})\right),
\end{eqnarray*}
where $\pi(\mu_q)=\tilde{q}\in N$.

Since $\gamma^* {\rm d}h$ is $\pi_N$-basic form, we have
\begin{equation*}
 {\rm T}^* {\rm i}_{{\mathcal L}_{N,\gamma}}\left(\gamma^* {\rm d}h(\tilde{q})\right)=\gamma^* {\rm d}h(\tilde{q})
\end{equation*}
and the result follows.
\qed
\end{prop}
Hence classical Hamilton-Jacobi equation in Theorem~\ref{thm:HJ} for a closed 1-form $\gamma$ on a submanifold $N$ of $Q$ is recovered as the extended Hamilton-Jacobi for the Lagrangian submanifold ${\mathcal L}_{N, \gamma}$ of $T^*Q$.

Locally, $\gamma={\rm d}S$ where $S$ is a local function on $N$ and $\phi^\alpha$ are constraints defining implicitly $N$.
Let $\tilde{S}$ be an arbitrary extension of $S$ to a neighborhood of $Q$, we have that
\[
{\mathcal L}_{N,\gamma}=\{ \mu\in T^*Q\; |\; \mu={\rm d}\tilde{S}+\lambda_\alpha {\rm d}\phi^\alpha\}\, .
\]
Therefore, the Hamilton-Jacobi equation ${\rm T}^* {\rm i}_{{\mathcal L}_{N,\gamma}}\left({\rm d}H\right)=0$ is equivalent to
\[
H({\rm d}\tilde{S}(q)+\lambda_\alpha {\rm d}\phi^\alpha(q))=\hbox{constant} \quad \forall \; q\in N.
\]
That is, we have the following system of equations
\begin{equation*}
\left.\begin{array}{rcl}
H\left(q^i, \dfrac{\partial \tilde{S}}{\partial q^i}+\lambda_\alpha\dfrac{\partial \phi^\alpha}{\partial q^i}\right)&=&\hbox{constant},\\[2mm]
\phi^\alpha(q^i)&=&0.
\end{array} \right\}
\end{equation*}

On the other hand, the submanifold $N$ can be locally immersed in $Q$ as follows ${\rm i}_N\colon (q^a)\hookrightarrow (q^a,\Psi^\alpha(q^a))$.
Then the constraints $\phi^\alpha$ become $\phi^\alpha(q)=q^\alpha-\Psi^\alpha(q^a)$ and  ${\rm T}^*{\rm i}_N\colon T^*_NQ \rightarrow T^*N$ is locally given by
\begin{equation*}
(q^a,p_i)\rightarrow \left(q^a,p_a+p_\alpha \dfrac{\partial \Psi^\alpha}{\partial q^a} \right).
\end{equation*}
From here, we have
\begin{eqnarray*}
H_{|{\mathcal L}_{N,\gamma}}(q^i,p_i)&=&H({q^i,\rm d}\tilde{S}(q)+\lambda_\alpha {\rm d}\phi^\alpha(q))\\
&=&H\left(q^i,\dfrac{\partial \tilde{S}}{\partial q^a}-\lambda_\alpha \dfrac{\partial \Psi^\alpha}{\partial q^a},\lambda_\alpha\right)\\ &=&(h\circ {\rm i}_N^*)\left(q^i,\dfrac{\partial \tilde{S}}{\partial q^a}-\lambda_\alpha \dfrac{\partial \Psi^\alpha}{\partial q^a},\lambda_\alpha \right)\\ &=& h\left(q^a, \dfrac{\partial {S}}{\partial q^a}\right).
\end{eqnarray*}
Thus we obtain the Hamilton-Jacobi equation on $N$, $h\circ {\rm d}S=\hbox{constant}$, from the Hamilton-Jacobi equation on $T^*Q$
as stated in Proposition \ref{holonomic}.
This approach is useful to understand dynamics on general manifolds: first we embed the manifold in an appropriate Euclidean space
by the Whitney immersion theorem~\cite{Lee} and then we solve the extended equations to obtain information about the solutions on the original manifold.

\subsection{Nonholonomic mechanics}\label{SubSec:NonholonomicNoTime}
Consider a mechanical system  specified by a regular lagrangian $L: TQ\longrightarrow \R$ and a submanifold $D$ of $TQ$ called the constraint submanifold.
If the submanifold $D$ is not of the form $TN$ for a submanifold $N$ of $Q$, the constraints are called nonholonomic.
We restrict ourselves to the case of linear constraints where $D$ is a  vector subbundle of $TQ$.

The regular Lagrangian defines a vector field $\Gamma_L$ on $T^*Q$
by the equation
\[
 {\rm i}_{\Gamma_L}\omega_L={\rm d}E_L,
\]
where $\omega_L=-{\rm d}\Theta_L$ is the Poincar\'e-Cartan 2-form where $\Theta_L=S^*({\rm d}L)$ and $E_L=\Delta L-L$ is the energy of the system.
The regularity condition of the Lagrangian $L$ is equivalent to the symplecticity of the 2-form $\omega_L$. Here, $\Delta$ is the Liouville vector field on $TQ$ and the canonical tensor field $S$ is called the vertical endomorphism. In local natural coordinates $(q^i, \dot{q}^i)$ on $TQ$ by
\[
\Delta=\dot{q}^i\frac{\partial}{\partial \dot{q}^i}, \quad S=\frac{\partial}{\partial \dot q^i}\otimes {\rm d}q^i.
\]
The vector field $\Gamma_L$ is a second order vector field (SODE), that is, it verifies $S(\Gamma_L)=\Delta$, but it is not in general tangent to $D$. That is why it is necessary to introduce forces to modify the dynamics in such a way that the solutions are integral curves of a vector field tangent to $D$. Let us assume that there is a set of reaction forces, mathematically expressed as a set $F\subset T^*_DTQ$. Therefore, the equations of the dynamics are now given by
\begin{equation}\label{asdf}
{\rm i}_{\Gamma_{nh}}\omega_L-{\rm d}E_L\in F,   \qquad \Gamma_{nh}\in \mathfrak{X}(D)\, ,
\end{equation}
where $ \mathfrak{X}(D)$ is the space of vector fields on $D$.

For nonholonomic systems the reaction forces are of ``mechanical type" (see \cite{Marle}) or, in other words,  they are given by semibasic 1-forms, that is,  $S^*(\mu)=0$ for all $\mu\in F$. Therefore, the possible solutions of (\ref{asdf}) automatically verify the SODE condition, that is, $S(\Gamma_{nh})=\Delta$.
A special choice for the reaction forces is given by Chetaev's condition which can be intrinsically given by
$F=S^*(TD^0)$.
 %
 Under this condition the system is conservative. In other words, the energy $E_L$ remains constant along the motion. From now on, we restrict ourselves to this case.

Denote by $W\subset T_DTQ$ the subbundle defined by $W^{0}=F$. In the sequel, we assume the following compatibility condition
\begin{equation}\label{zaq}
T_D(TQ)=TD\oplus W^{\perp}\; ,
\end{equation}
where  $W^{\perp}$ denotes the symplectic orthogonal to $W$ with respect to the symplectic form $\omega_L$. Observe that by construction $W^{\perp}$  is contained in the vertical tangent bundle to $TQ$ on $D$.

Condition (\ref{zaq}) automatically implies the existence of a unique solution $\Gamma_{nh}$ of Equations (\ref{asdf}).
Now let us assume that the Lagrangian is hyper-regular, that is, the Legendre transformation $\F L: TQ\longrightarrow T^*Q$, locally defined by
\[
\F L(q^i, \dot{q}^i)=\left(q^i, p_i=\dfrac{\partial L}{\partial \dot{q}^i}\right)\; ,
\]
 is a diffeomorphism.
 Under this assumption we have an equivalent hamiltonian formulation for the nonholonomic mechanics.Then the new ingredients are a hamiltonian $H: T^*Q\rightarrow \R$ defined by $H=E_L\circ (\F L)^{-1}$,  the constraint submanifold given by ${C}=\F L (D)$ and the vector subbundle $\tilde{F}\in T_{C}T^*Q$ defined by $(\F L)^*\tilde{F}=F$. We also have
$\tilde{W}=  (\F L)_* W$, then the compatibility condition~\eqref{zaq} is rewritten as
\[
T_{C}(T^*Q)=T{C}\oplus \tilde{W}^{\perp}\; ,
\]
where  $\tilde{W}^{\perp}$ as above denotes the symplectic orthogonal to $\tilde{W}$ with respect to the canonical symplectic form ${\omega_Q}$.
Moreover, $\tilde{W}^{\perp}$ is contained in the vertical tangent bundle to $T^*Q$. Consequently, $\tilde{W}^{\perp}$ is isotropic. Thus,
\[
\tilde{W}=\tilde{W}\cap \left(T{C}\oplus \tilde{W}^{\perp}\right)=\left(\tilde{W}\cap T{C}\right)\oplus \tilde{W}^{\perp}\; ,
\]
and we conclude that ${\mathcal H}=\tilde{W}\cap T{C}$ is a symplectic vector bundle on $C$, that is,
\[
{\mathcal H}\oplus {\mathcal H}^{\perp}=T_{{C}}T^*Q\; .
\]
The  dynamics on the Hamiltonian formalism is given by the following equations:
\begin{equation*}\label{asdf-1}
{\rm i}_{\xi_{nh}}{\omega_Q}-{\rm d} H\in \tilde{W}^0,   \qquad \xi_{nh}\in TC\; .
\end{equation*}
Moreover, as $D$ is a vector subbundle, it is not hard to prove that
\[
\xi_{nh}(x)\in {\mathcal H}_x, \quad \forall \quad x\in C\; .
\]
In other words, $\xi_{nh}(x)\in \tilde{W}_x$ for every $x\in C$.


\begin{prop}\label{aqwe} Let ${\mathcal L}$ be a  Lagrangian distribution on the symplectic vector subbundle ${\mathcal H}$ on $C$. The following statements are equivalent:
\begin{enumerate}
\item $\langle {\rm d}H(y), {\mathcal L}_y\rangle =0$, that is, $\langle {\rm d}H(y), X(y)\rangle =0$ for every section $X$ of ${\mathcal L}$,
\item $(\xi_{nh})(y)\in {\mathcal L}_y$,
\end{enumerate}
for every $y \in C$.
\proof
Since ${\mathcal L}\subset {\mathcal H}$ we have that  $\langle {\rm d}H, {\mathcal L}\rangle=0$  is equivalent to
 $\omega_Q(\xi_{nh}, {\mathcal L})=0$. Therefore $\xi_{nh}(y)\in {\mathcal L}_y^{\perp}={\mathcal L}_y$  for every $y\in C$
because ${\mathcal L}$ is a Lagrangian distribution where 
\[
{\mathcal L}^\perp=\{ V\in{\mathcal H}\; |\; {\omega_Q}(V, {\mathcal L})=0\}\; .
\]. 
\qed
\end{prop}


Moreover, we have that ${\rm T}\pi_Q({\mathcal H})=D$ and if $\gamma$ is a section of the fiber bundle
$(\pi_Q)_{|C}: C\rightarrow Q$, then ${\rm T}\gamma(D)\in {\mathcal H}$.
Thus, if we define the vector field $\xi_{nh}^{\gamma}$ on $Q$ by
\[
\xi_{nh}^{\gamma}(x)={\rm T}_{\gamma(x)}\pi_Q(\xi_{nh}(\gamma(x))),
\]
then we have the following corollary, as  a particular case of Proposition \ref{aqwe}. (See \cite{Iglesias,AlmostPoisson})
\begin{corol}\label{aqwe-1} If ${\rm T}\gamma(D)$ is  a  Lagrangian distribution on ${\mathcal H}$ along the points of $\gamma(Q)$. The following statements are equivalent:
\begin{enumerate}
\item $\langle {\rm d}(H\circ \gamma), D\rangle =0$,
\item $\xi_{nh}$ and $\xi_{nh}^{\gamma}$ are $\gamma$-related.
\end{enumerate}
\proof
The proof follows the same lines as Proposition \ref{aqwe}. \qed
\end{corol}

Observe that if ${\rm T}\gamma(D)$ is a Lagrangian distribution along $\gamma(Q)$, then it implies that for all $v_1, v_2\in D$ we have  that
\begin{eqnarray*}
0&=&{\omega_Q}({\rm T}\gamma(v_1), {\rm T}\gamma (v_2))\\
 &=& -(\gamma^*{\rm d}\theta)(v_1,v_2)\\
  &=&-{\rm d}(\gamma^*\theta)(v_1, v_2)\\
  &=&-{\rm d}\gamma(v_1, v_2)
  \end{eqnarray*}
where we have used the fact that $\gamma^*\theta=\gamma$, see Section~\ref{Sec:Lagrang} for more details. This is precisely the condition that appears in previous approaches to Hamilton-Jacobi theory for nonholonomic systems, see~\cite{Iglesias,AlmostPoisson,TomokiB,OFBZ}.

\remark If $D$ is completely nonholonomic, that is, the smallest involutive distribution under the Lie bracket which contains $D$
is the entire tangent bundle, then the condition (i) in Proposition~\ref{aqwe} implies that $H\circ \gamma$ is constant. This last condition is analogous to condition (ii) in Theorem~\ref{thm:HJ}.

\section{Generalized time-dependent Hamilton-Jacobi theory}\label{Sec:TimeHJ}

From now
on, we focus on the geometric version of time-dependent Hamilton-Jacobi equation in terms of Lagrangian submanifolds.
We first state a general Hamilton-Jacobi theorem for a time-dependent 1-form and a family of  Lagrangian submanifolds of $(T^*Q,{\omega_Q})$,  extending Section~\ref{Sec:GHJParticular} to the non-autonomous case.


Let   $\alpha_t$ be a time-dependent family of 1-forms on $T^*Q$ and  ${\mathcal L}_t$ be Lagrangian submanifolds  of $(T^*Q,{\omega_Q})$, we define the following family of affine distributions on $TT^*Q$ fibering over ${\mathcal L}_t$
\begin{equation} {\Sigma}_t=\sharp\left(\left.\alpha_t\right|_{{\mathcal L}_t} \right)  + T {\mathcal L}_t, \label{eq:StGeneral}
\end{equation}
where $\sharp\colon T^*T^*Q\rightarrow TT^*Q$ is the musical isomorphism.

With these elements we obtain a more general  version of time-dependent Hamilton-Jacobi result than the one stated in
Theorem~\ref{thm:HJtime}.

\begin{thm} Let ${\mathcal L}_t$ be a family of Lagrangian submanifolds of $(T^*Q,{\omega_Q})$, $\alpha_t$ be a family of 1-forms on $T^*Q$, $({\Sigma}_t)_{t\in \R}$ be defined as in~\eqref{eq:StGeneral} and
$H$ be a Hamiltonian function on $T^*Q$.
The following statements are equivalent:
\begin{enumerate}
\item \label{Cond1TimeG}
${\rm T}^*{\rm  i}_{{\mathcal L}_t}\left({\rm d} H- \alpha_t\right)=0$ for every $t\in \mathbb{R}$, where ${\rm i}_{{\mathcal L}_t}\colon {\mathcal L}_t \rightarrow T^*Q$ is the canonical immersion,
\item \label{Cond2TimeG}
$ \left. \left({\rm d}H-
\alpha_t\right)\right|_{{\mathcal L}_t} \in  (T{\mathcal L}_t)^0
$,
\item \label{Cond3TimeG}
$\left. X_H\right|_{{\mathcal L}_t}\in \Sigma_t$.
\end{enumerate}

\proof
Note that condition~\ref{Cond1TimeG} is straightforward equivalent to condition~\ref{Cond2TimeG}. By applying the sharp musical isomorphism to~\ref{Cond2TimeG}, $({\rm d}H)_{|{\mathcal L}_t}-
\alpha_t\in  (T{\mathcal L}_t)^0$, we have
\begin{equation*}
\sharp\left( \left. \left({\rm d}H-
\alpha_t\right)\right|_{{\mathcal L}_t}\right)\in \sharp\left( (T{\mathcal L}_t)^0\right).\end{equation*}
Equivalently, we have
\begin{equation*}
\sharp\left(  \left({\rm d}H-
\alpha_t\right)_{p_q}\right)\in \sharp\left( (T_{p_q}{\mathcal L}_t)^0\right)=\left(T_{p_q}{\mathcal L}_t\right)^\perp=T_{p_q}{\mathcal L}_t,
\end{equation*}
for every $p_q\in {\mathcal L}_t$,
 because ${\mathcal L}_t$ is a Lagrangian submanifold. Thus,
\begin{equation*}
\left. X_H\right|_{{\mathcal L}_t}\in \sharp\left(\left.\alpha_t\right|_{{\mathcal L}_t}\right)+T{\mathcal L}_t=\Sigma_t
\end{equation*}
and this concludes the proof because all the arguments can be reversed.
\qed \label{Thm:TimeHJGen}
\end{thm}

Now let us consider the particular case where the family of Lagrangian submanifolds ${\mathcal L}_t$ is given by a time-dependent
family of  closed 1-forms on $Q$.
In other words, consider a map $\gamma\colon \mathbb{R}\times Q \rightarrow T^*Q$
such that $\gamma_t\colon Q \rightarrow T^*Q$ is a closed 1-form
on $Q$ for all $t\in \mathbb{R}$.
 Locally, $\gamma (t,q)=\gamma_t(q)=(q^i,\gamma_i(t,q))$.

For each $t$, define the subset ${\mathcal L}_t={\rm Im} \gamma_t \subset
T^*Q$. As in the general case at the beginning of this section, we cannot only consider the tangent space to ${\mathcal L}_t$ as  in Section 4 because of the
time-dependence of the family of sections. That is why
we define the following family of affine distributions on  $TT^*Q$ fibering over  ${\mathcal L}_t=\hbox{Im}\gamma_t$:
\begin{equation} {\Sigma}_t={\rm T}\gamma\left(\frac{\partial}{\partial t}\right) + {\rm T}\gamma_t (TQ), \label{eq:St}
\end{equation}
where ${\rm T} \gamma \colon T(\mathbb{R}\times Q)\rightarrow TT^*Q$ and ${\rm T} \gamma_t\colon TQ\rightarrow TT^*Q$.
Locally,
\begin{equation*} {\Sigma}_t=\left\{\left(q^i,\gamma_i(t,q), \dot{q}^i, \frac{\partial \gamma_i}{\partial t}(t,q)+\dot{q}^j\frac{\partial \gamma_i}{\partial q^j}(t,q)\right)\in TT^*Q\; | \; \forall \; (q, \dot{q})\in TQ
\right\}.
\label{eq:StLocal}
\end{equation*}
In other words,
\begin{equation*}
 ({\Sigma}_t)_{\gamma_t(q)}=\frac{\partial \gamma_i}{\partial
t}\,\frac{\partial}{\partial p_i}_{\gamma_t(q)}+{\rm
span}_{\mathbb{R}}\left\{\frac{\partial}{\partial
q^j}+\frac{\partial \gamma_i}{\partial q^j}\,
\frac{\partial}{\partial p_i}\right\}_{\gamma_t(q)}.
\end{equation*}
Consider the 1-form $\alpha_t$ on $T^*Q$ defined by
\[
\alpha_t=\flat \left({\rm T}\gamma\left(\frac{\partial}{\partial t}\right)\right)\, ,
\]
with local expression $\alpha_t(q,\gamma_i(t,q))=-\dfrac{\partial \gamma_i}{\partial t} (t,q) \, {\rm d}q^i$.
Now we can state the following corollary as a particular case of Theorem~\ref{Thm:TimeHJGen}.
\begin{corol} Let ${\mathcal L}_t={\rm Im} \gamma_t \subset
T^*Q$ where $\gamma_t$ is a family of closed 1-forms on $Q$ and $({\Sigma}_t)_{t\in \R}$ be defined as in~\eqref{eq:St}. Define the family of vector fields $X_H^{\gamma_t}\in {\mathfrak X}(Q)$
by
\[
X_H^{\gamma_t}= {\rm T}\pi_Q\circ X_H\circ \gamma_t\, .
\]
The following statements are equivalent:
\begin{enumerate}
\item \label{Cond4Time}
${\rm d}(H\circ \gamma_t)+ \dfrac{\partial
\gamma_i}{\partial t} \, {\rm d} q^i=0$;
\item \label{Cond3Time}
$ {\rm d}(H)_{|{\mathcal L}_t}-
\alpha_t\in  (T{\mathcal L}_t)^0
$;
\item \label{Cond1Time}
$\left.X_H\right|_{{\mathcal L}_t}\in \Sigma_t$;
\item \label{Cond2Time}
if $c: I\rightarrow Q$ is an integral curve of $X_H^{\gamma_t}$, then $\gamma_t\circ c: I\rightarrow T^*Q$
is an integral curve of $X_H$.
\end{enumerate}

\proof \ref{Cond1Time} $\Leftrightarrow$~\ref{Cond2Time}. Locally,
 $(X_H)_{|{\mathcal L}_t}\in \Sigma_t$ is equivalent to
 \begin{equation}\label{moi}
 -\dfrac{\partial H}{\partial q^i}=\dfrac{\partial \gamma_i}{\partial t}+ \dfrac{\partial \gamma_i}{\partial q^j}\dfrac{\partial H}{\partial p_j}
\end{equation}
along $\hbox{Im }\gamma_t$.
To prove \ref{Cond2Time} we  start with an integral curve $c: I\rightarrow Q$ of $X_H^{\gamma_t}$. If locally $c(t)=(q^i(t))$, then \[
\dfrac{{\rm d} q^i}{{\rm d}t}=\frac{\partial H}{\partial p_i}(q^i(t), \gamma_i(t,q(t)))\; .
\]
Now, using  (\ref{moi}) we have that
\begin{eqnarray*}
\frac{{\rm d} \left(\gamma_i\circ c\right) }{{\rm d}t}&=&\frac{\partial\gamma_i}{\partial t}+\frac{\partial \gamma_i}{\partial q^j}\frac{\partial H}{\partial p_j}\\
                           &=& -\frac{\partial H}{\partial q^i}\; .
                            \end{eqnarray*}
Therefore, $\gamma_t\circ c: I\rightarrow T^*Q$
is an integral curve of $X_H$.
The converse can be proved by  reversing the arguments.

The remaining equivalences follow straightforward from Theorem~\ref{Thm:TimeHJGen}.

%
%
%
%
\qed
\label{thm:TimeHJPart}
\end{corol}

\remark
Note that for a function $W\colon \mathbb{R}\times Q\rightarrow
\mathbb{R}$, we have
$$\xymatrix{ \mathbb{R}\times Q \ar[r]^{\txt{\small{${\rm d}W$}}} \ar[rd]^{\txt{\small{$\gamma$}}}&
T^*(\mathbb{R}\times Q)\ar[d]^{\txt{\small{$\pi_2$}}}  \\
& T^*Q }$$
Then in  Corollary~\ref{thm:TimeHJPart} we have that
\begin{equation*}
 \gamma_i(t,q){\rm d}q^i=\dfrac{\partial W}{\partial q^i}(t,q){\rm d}q^i.
\end{equation*}
Thus condition (i) in Corollary~\ref{thm:TimeHJPart} can be rewritten as
\begin{equation*}
 {\rm d}\left(H\circ \gamma_t-\dfrac{\partial W}{\partial t}\right)=0,
\end{equation*}
or equivalently
\begin{equation*}
H\left(q^i, \frac{\partial W}{\partial q^i}\right)+\frac{\partial W}{\partial t}=K(t)
\end{equation*}
where $K(t)\in \R$,
which is  the same condition as the one in (ii) in Theorem~\ref{thm:HJtime}.

\section{Applications of time-dependent Hamilton-Jacobi theory}\label{SubSecTimeHJ}

Analogously to Section~\ref{SubSec} we use here particular time-dependent families of the Lagrangian submanifolds in~\eqref{eq:StGeneral} to  develop the novel applications of Theorem~\ref{Thm:TimeHJGen} in non-autonomous Hamilton-Jacobi theory.

\subsection{Holonomic time-dependent dynamics} Let $N$ be a submanifold of $Q$ and $h$ be a Hamiltonian function on $T^*N$.
Take $\gamma\colon \mathbb{R}\times N \rightarrow T^*N$ such that $\left\{ \gamma_t\right\}_{t\in \mathbb{R}}$ is a family of
closed 1-forms $\gamma_t$ on $N$.
Then, we have that ${\mathcal L}_t=\hbox{Im}\, \gamma_t$ is a Lagrangian submanifold of $T^*N$ and at the same time
we can define another family of Lagrangian submanifolds  ${\mathcal L}_{N,\gamma_t}$ of $T^*Q$ according to~\eqref{eq:Sigma_mf}
which is given by
\begin{equation*}
{\mathcal L}_{N,\gamma_t}=\{\mu\in T^*Q \; | \; {\rm T}^*{\rm i}_N(\mu)=\gamma_t\}=\{\mu \in T^*Q \; | \; \mu_{{\rm i}_{N}(x)}({\rm T}_x{\rm i}_{N}(v_x))=\gamma_{t,x}(v_x) \; \forall \; v_x\in T_xN \},
\end{equation*}
where ${\rm i}_{N} \colon N \hookrightarrow Q$ is the canonical immersion.
This situation is summarized in the following diagram:
\begin{equation*}
 \xymatrix{ \mathbb{R} & T^*N \ar@/^1pc/[dr]_{\txt{\small{$\pi_{N}$}}} \ar[l]_{\txt{\small{$h$}}} &
T^*Q \ar[d]^{\txt{\small{$\pi_Q$}}} \ar[l]_{\txt{\small{${\rm T}^*{\rm i}_{N}$}}}  & {\mathcal L}_{N,\gamma_t}
 \ar@{_{(}->}[l]_{\txt{\small{${\rm i}_{{\mathcal L}_{N,\gamma_t}}$}}} \ar[d]^{\txt{\small{$\left.\pi_Q\right|_{{\mathcal L}_{N,\gamma_t} }$}}} \\ & & N  \ar@/^1pc/[ul]^{\txt{\small{$\gamma_t$}}} \ar@{=}[r]  & N
}
\end{equation*}
for every $t\in \mathbb{R}$.
Hence we can extend the Hamiltonian $h$ to a hamiltonian $H: T^*_{N} Q\longrightarrow {\mathbb R}$ defined by $H= h\circ {\rm T}^*{\rm i}_{N}$.

As described at the beginning of Section~\ref{Sec:TimeHJ}, the time-dependence in the dynamics forces us to introduce a 1-form
$\alpha_t$ on $T^*N$. In order to describe the holonomic time-dependent Hamilton-Jacobi equation on the manifold $T^*_NQ$ we must
define a family of 1-forms $\Pi_t$ on $T^*_NQ$ obtained from $\alpha_t$ as follows:
\begin{equation}
 \langle \Pi_t(\mu_q),X(\mu_q)\rangle=\langle \alpha_t, {\rm  T}_{\mu_q}\left({\rm T}^* {\rm i}_N\right) X(\mu_q) \rangle,
\label{eq:Pit}
\end{equation}
for all $\mu_q\in T^*_NQ$, $X(\mu_q)\in T_{\mu_q}T^*_NQ$, ${\rm  T}_{\mu_q}\left({\rm  T}^* {\rm i}_N\right)\colon T_{\mu_q}T^*_NQ \rightarrow
T_qT^*N$. In fact, $\Pi_t={\rm  T}^*({\rm  T}^*{\rm i}_N)\left( \alpha_t \right)$ because ${\rm  T}^*({\rm  T}^*{\rm i}_N)\colon T^*(T^*N)\rightarrow T^*(T^*_NQ)$.

\begin{prop} Let $\alpha_t$ be a time-dependent family of 1-forms on $T^*N$, standard time-dependent Hamilton-Jacobi equation for holonomic dynamics on $N$
\[ \gamma_t^* {\rm d}h-{\rm  T}^*{\rm i}_{{\mathcal L}_t}(\alpha_t)=0, \mbox{ that is, }
h\circ \gamma_t-\left.\alpha_t\right|_{{\mathcal L}_t}= {\rm constant } \mbox{ for every } t,
\]
is equivalent to
\[
{\rm T}^*{\rm i}_{{\mathcal L}_{N,\gamma_t}}({\rm d}H-\Pi_t)=0,
\]
where $H: T^*_{N} Q\longrightarrow {\mathbb R}$, $H= h\circ {\rm T}^*{\rm i}_{N}$ and $\Pi_t$ is the 1-form on $T^*_NQ$ defined in~\eqref{eq:Pit}.
\proof
For $\mu_q\in {\mathcal L}_{N,\gamma_t} \subseteq T^*_NQ$,
\begin{eqnarray*}
0&=& \langle {\rm T}^*{\rm i}_{{\mathcal L}_{N,\gamma_t}}({\rm d}H-\Pi_t), X_{\mu_q}\rangle =
\langle ({\rm d}H-\Pi_t)_{\mu_q}, {\rm T}_{\mu_q}{\rm i}_{{\mathcal L}_{N,\gamma_t}} X_{\mu_q}\rangle\\
&=& \langle ({\rm d} \left(h\circ {\rm T}^*{\rm i}_{N}\right)-{\rm T}^*({\rm T}^*{\rm i}_N)\left( \alpha_t \right))_{\mu_q}, {\rm T}_{\mu_q}{\rm i}_{{\mathcal L}_{N,\gamma_t}} X_{\mu_q}\rangle\\
&=& \langle {\rm d} \left(h\circ \gamma_t\right)_{\mu_q}, {\rm T}_{\mu_q}{\rm i}_{{\mathcal L}_{N,\gamma_t}} X_{\mu_q}\rangle-\langle {\rm T}^*({\rm T}^*{\rm i}_N)\left( \alpha_t \right)_{\mu_q}, {\rm T}_{\mu_q}{\rm i}_{{\mathcal L}_{N,\gamma_t}} X_{\mu_q}\rangle\\
&=& \langle {\rm d} \left(h\circ \gamma_t\right)_{\mu_q}, {\rm T}_{\mu_q}\left.(\pi_Q)\right|_{{\mathcal L}_{N,\gamma_t}} X_{\mu_q}\rangle-\langle \left( \alpha_t \right)_{\mu_q}, {\rm T}_{ {\rm i}_{{\mathcal L}_{N,\gamma_t}}(\mu_q) }({\rm T}^*{\rm i}_N) {\rm T}_{\mu_q}{\rm i}_{{\mathcal L}_{N,\gamma_t}} X_{\mu_q}\rangle\\
&=& \langle {\rm d} \left(h\circ \gamma_t\right)_{\mu_q}, {\rm T}_{\mu_q}\left.(\pi_Q)\right|_{{\mathcal L}_{N,\gamma_t}} X_{\mu_q}\rangle-\langle \left( \alpha_t \right)_{\mu_q}, {\rm T}_q{\rm i}_{{\mathcal L}_t} {\rm T}_{\mu_q}\left.(\pi_Q)\right|_{{\mathcal L}_{N,\gamma_t}} X_{\mu_q}\rangle.
\\&=&\langle \left({\rm d} \left(h\circ \gamma_t\right)-{\rm T}^*{\rm i}_{{\mathcal L}_t}(\alpha_t)\right)_{\mu_q}, {\rm T}_{\mu_q}\left.(\pi_Q)\right|_{{\mathcal L}_{N,\gamma_t}} X_{\mu_q}\rangle \, .
\end{eqnarray*}
The above equalities are proved using the same arguments as in the proof of Proposition~\ref{holonomic}.
From here the result follows.
\qed
\label{thm:HJHolonomicTime}
\end{prop}

Locally, $\gamma_t={\rm d}W_t$ with $W: \R\times N\rightarrow \R$  and $\phi^\alpha$ are constraints defining implicitly $N$. Consider an arbitrary extension $\tilde{W}$   of $W$ to a neighborhood of $\R\times Q$ so that
\[
{\mathcal L}_{N,\gamma_t}=\{ \mu\in T^*Q\; |\; \mu={\rm d}\tilde{W}_t+\lambda_{t, \alpha} {\rm d}\phi^\alpha\}\, .
\]
Therefore, the Hamilton-Jacobi equation  is rewritten as
\begin{equation*}
\left.\begin{array}{rcl}
H({\rm d}\tilde{W}(q)+\lambda_{t, \alpha} {\rm d}\phi^\alpha(q)) + \dfrac{\partial \tilde{W}}{\partial t}& = &k(t),\\[3mm]
\phi^\alpha(q^i)&=&0.
\end{array} \right\}
\end{equation*}

\subsection{Nonholonomic time-dependent mechanics} Having in mind the description of nonholonomic mechanics in Section~\ref{SubSec} the time-dependent case works as follows. Consider a family of Lagrangian distributions ${\mathcal L}_t$ on the symplectic vector subbundle ${\mathcal H}$
on $C$ and a family of 1-forms $\alpha_t$ on $T^*Q$. Instead of considering the tangent bundle ${\rm T} {\mathcal L}_t$ we must define the following submanifold
of $T_CT^*Q$:
\begin{equation}
 \Sigma_t=\sharp\left( \left.\alpha_t\right|_C \right)+{\rm T} {\mathcal L}_t.
\label{eq:StGeneralTime}
\end{equation}

\begin{prop} Let ${\mathcal L}_t$ be a family of Lagrangian distributions  on the symplectic vector subbundle ${\mathcal H}$
on $C$, $\alpha_t$ be a family of 1-forms on $T^*Q$ and $({\Sigma}_t)_{t\in \R}$ be defined as in~\eqref{eq:StGeneralTime}. The
following statements are equivalent:
\begin{enumerate}
\item $\langle \left({\rm d}H-\alpha_t\right)(y), \left({\mathcal L}_t\right)_y\rangle =0$, that is, $\langle {\rm d}H(y), X(y)\rangle =0$ for every section $X$ of ${\mathcal L}_t$,
\item $(\xi_{nh})(y)\in \left(\Sigma_t\right)_y$,
\end{enumerate}
for every $y \in C$ and for every time $t$. \label{thm:NonHTimeDepHJ}
\end{prop}

\remark As in Subsection~\ref{SubSec:NonholonomicNoTime}, for the time-dependent nonholonomic case it is also necessary to adapt the theory developed in Section~\ref{Sec:TimeHJ} to the framework of
symplectic vector subbundles by using Lagrangian distributions instead of Lagrangian submanifolds.

Now we consider the particular case of Lagrangian distributions on $H$ defined by a family of sections $\gamma_t$ of the fiber bundle
$\left.(\pi_Q)\right|_C\colon C \rightarrow Q$. These sections satisfy ${\rm T} \gamma_t (D) \in {\mathcal H}$. For this particular
case the
submanifold $\Sigma_t$ in~\eqref{eq:StGeneralTime} becomes
\begin{equation*}
 \Sigma_t={\rm T}\gamma \left( \dfrac{\partial}{\partial t}\right)+{\rm T} \gamma_t (D).
\end{equation*}
If we also define the
vector field $\xi_{nh}^{\gamma_t}$ on $Q$ by
\[
\xi_{nh}^{\gamma_t}(x)={\rm T}_{\gamma_t(x)}\pi_Q(\xi_{nh}(\gamma_t(x))),
\]
then we have the following corollary  as  a particular case of Proposition \ref{thm:NonHTimeDepHJ}.
\begin{corol}\label{thm:NonHTimeDepHJParticular} If ${\rm T}\gamma_t(D)$ is  a  Lagrangian distribution on ${\mathcal H}$ along the points of $\gamma_t(Q)$. The following statements are equivalent:
\begin{enumerate}
\item $\left\langle {\rm d}(H\circ \gamma_t)+\dfrac{\partial \gamma_i}{\partial t} {\rm d}q^i, D\right\rangle =0$,
\item $\left.\xi_{nh}\right|_{{\rm T}\gamma_t(D)}\subseteq \Sigma_t$,
\item $\xi_{nh}$ and $\xi_{nh}^{\gamma_t}$ are $\gamma_t$-related.
\end{enumerate}
\proof
The proof follows the same lines as the ones of Proposition~\ref{aqwe} and Corollary~\ref{thm:TimeHJPart}. \qed
\end{corol}

\section{Conclusions and future developments}

In this paper we have described a general version of Hamilton-Jacobi equation which allows us to consider different examples of mechanical systems such as  holonomic, nonholonomic  or time-dependent systems  within the same framework.
Our approach is based on the notion of Lagrangian submanifolds  as a tool to  prove a more general version of the Hamilton-Jacobi equation and to give special emphasis on applications. In particular, we derive a Hamilton-Jacobi equation for holonomic systems in such a way that the associated Lagrange multipliers are provided with a geometrical interpretation. Moreover, using the ``symplectic  approach" to nonholonomic mechanics (see \cite{BaSn,Marle}) we show that our techniques are also adapted to the nonholonomic version of Hamilton-Jacobi equation. This result is more general than those known in the literature. Finally, using time-dependent families of Lagrangian submanifolds we directly obtain the corresponding  versions of the time-dependent Hamilton-Jacobi equation.

One of the main advantages of this paper is the possibility to use the same version of the Hamilton-Jacobi equation to deal with very different situations such as time-dependency, presence of constraints, etc. Moreover, it is possible to adapt this framework to other cases like Hamilton-Jacobi equation for singular Lagrangian systems, reduced systems, systems whose dynamics is described by Poisson or almost-Poisson brackets and, even for classical field theories. It is also envisaged that the approach in this paper would contribute to get new insights into the study of caustics of Lagrangian submanifolds~\cite{1990Arnold,2012ArnoldGV}.


\section*{Acknowledgements}

This work has been partially supported by MICINN (Spain)
 MTM2008-00689, MTM2010-21186-C02-01 and MTM2010-21186-C02-02; 2009SGR1338 from the Catalan government; the European project IRSES-project “GeoMech-246981” and the ICMAT Severo Ochoa project SEV-2011-0087.
MBL has been financially supported by Juan de
la Cierva fellowship from MICINN.

\bibliographystyle{plain}
\bibliography{References-hj}
\end{document}